\pdfoutput=1
\documentclass[12pt]{iopart}
\usepackage{graphicx,units}
\usepackage{amssymb}

\usepackage{xr}
\begin{document}
	\title[]{Spatial inhomogeneity and temporal dynamics of a 2D electron gas in interaction with a 2D adatom gas}
	
	\author{F. Cheynis$^{1}$, S. Curiotto$^{1}$, F. Leroy$^{1}$, P. M\"{u}ller$^{1}$}
	\address{$^1$ Aix Marseille Univ, CNRS, CINAM, Marseille, France }
	
	\ead{cheynis@cinam.univ-mrs.fr}
	\vspace{10pt}
	\begin{indented}
		\item[] \today
	\end{indented}

%\affil[+]{these authors contributed equally to this work}

%\keywords{Keyword1, Keyword2, Keyword3}

\begin{abstract}
Fundamental interest for 2D electron gas (2DEG) systems has been recently renewed with the advent of 2D materials and their potential high-impact applications in optoelectronics. Here,
we investigate a 2DEG created by the electron transfer from a Ag adatom gas deposited on a Si(111)$\sqrt{3}\times\sqrt{3}$-Ag surface to an electronic surface state. Using low-energy electron microscopy (LEEM), we measure the Ag adatom gas concentration and the 2DEG-induced charge transfer.  %via the surface work function determination 
We demonstrate a linear dependence of the surface work function change on the Ag adatom gas concentration. A breakdown of the linear relationship is induced by the occurrence of the Ag adatom gas superstructure identified as Si(111)$\sqrt{21}\times\sqrt{21}$-Ag only observed below room temperature. We evidence below room temperature a confinement of the 2DEG on atomic terraces characterised by spatial inhomogeneities of the 2DEG-induced charge transfer along with temporal fluctuations. These variations mirror the Ag adatom gas concentration changes induced by the growth of 3D Ag islands and the occurrence of an Ehrlich-Schwoebel diffusion barrier of \unit[155$\pm$10]{meV}.
\end{abstract}

\flushbottom
\maketitle
\thispagestyle{empty}

%\noindent Please note: Abbreviations should be introduced at the first mention in the main text ? no abbreviations lists. Suggested structure of main text (not enforced) is provided below.

\section*{Introduction}

Owing to their fascinating properties, 2D electron gas systems (2DEG) have recently generated major breakthroughs in the field of condensed matter physics. For instance, the occurrence of a conductive 2D phase at the surface of oxides paves the way for the emerging field of functional oxide electronics \cite{san11}. 2D Dirac electron gas observed in graphene has been shown to surpass the long-standing 2DEG confined at the GaAs/AlGaAs interface to determine the Planck constant \emph{h} using Quantum Hall resistance measurements \cite{art-CHE15d}. %This allows for a redefinition of the \emph{Syst\`eme International d\textquoteright unit\'es} (SI) based on the relation between the kilogram standard and the Planck constant \emph{h}. 
2DEG at metal/semiconductor interfaces have pushed the phenomenon of superconductivity downto its very 2D limit \cite{zha10} and also shown intriguing electron localisation and metal-insulator transition \cite{mat07}. Apart from homogeneous 2DEG systems, a growing need for nanostructured electron gas \cite{her11,pal15,bha16} is motivated by accessible breakthroughs in low dimensional electronics.  %. For instance 1D channels obtained by lithographic approaches are building blocks for single-electron meso-transport studies \cite{her11}. 2DEG under an applied magnetic field can be localized at the expense of demanding experimental conditions ($T\leq\unit[1]{K}$ and $B_{\mathrm{appl.}}\geq\unit[10]{T}$) and only at the nanometer scale \cite{bha16}. %To the best of our knowledge, the occurrence of a self-structured 2DEG has not been reported so far.
Here, we focus on a well-documented 2DEG that is interestingly created by a charge transfer from a Ag 2D adatom gas (Ag-2DAG) to the so-called \emph{S$_\mathrm{1}$} electronic surface state of the Si(111)$\sqrt{3}\times\sqrt{3}$-Ag reconstructed surface ($\sqrt{3}$-Ag) \cite{nak97,aiz99,uhr02,cra05,liu06}. % More particularly, the 2DAG obtained by Ag deposition (Ag-2DAG) onto a Si(111)$\sqrt{3}\times\sqrt{3}$-Ag reconstructed surface ($\sqrt{3}$-Ag) shows an electron doping property of the so-called \emph{S$_\mathrm{1}$} surface state \cite{aiz99}% that has been clearly demonstrated by the surface conductivity increase 
This charge transfer is responsible for an increase of the surface electrical conductance of \unit[50]{\%} at room-temperature (RT) upon a Ag deposition as small as \unit[0.03]{ML} on the $\sqrt{3}$-Ag surface\cite{nak97}. %The electron filling of the surface state is limited by the nucleation of a Ag 3D growing phase. %The system deviates from a pure free-electron-like dispersion relationship as the electron doping is increased owing to a surface state hybridization % with an impurity/atom localized level at LT 
%\cite{cra05,liu06}%and inter-surface state interactions at RT \cite{liu06}
% tocharacterize independently both atomic and electronic components of the system.
 
In this paper, we evidence the unreported mutual confinement of a 2DEG and a 2DAG on atomic terraces as large as a few $\mu m^2$ in the temperature range \unit[210-250]{K} that also evolves in time. More specifically we perform measurements of the Ag-2DAG concentration, $c$, and of the 2DEG-induced surface work function change using a single mesoscopic microscopy technique (low-energy electron microscopy, LEEM, see Methods for details). To validate our fine comprehension of the system, we first focus on the transition between the $\sqrt{3}$-Ag and the Si(111)$\sqrt{21}\times\sqrt{21}$-Ag surface reconstruction ($\sqrt{21}$-Ag) only observed below RT \cite{zha95}. The demonstration of a linear dependence of the surface work function change on the Ag-2DAG concentration along with results from the literature confirm that the electron doping of the 2DEG is revealed by the work function changes. 
%In this report, we demonstrate a linear relationship between the surface work function and the Ag-2DAG concentration, $c$, using a single mesoscopic microscopy technique (low-energy electron microscopy, LEEM). Comparison with results from the literature confirms that the electron filling of the 2DEG is revealed by the work function changes. 
%we first characterise the Ag-2DAG, as a preliminary step before a detailed analysis of the 2DEG. %In particular we will determine energy barriers for atomic processes such as adatom formation and diffusion by monitoring the concentration of the Ag-2DAG in interaction with a 3D-Ag phase. 
%In a second part, 
A breakdown of this linear relationship is observed above a critical Ag-2DAG concentration. %below RT %a non-trivial relationship between the surface work function and the Ag-2DAG concentration is evidenced
 Using a simple analytic model, we conclude that the breakdown results from the occurrence of the $\sqrt{21}$-Ag reconstruction. Below RT, we evidence the confinement of the 2DEG on atomic terraces upon Ag deposition and after the nucleation of 3D growing Ag islands. This regime is characterised by inhomogeneous spatial distributions and temporal fluctuations of the 2DEG charge transfer induced by the Ag-2DAG. %These two quantities recover a linear relationship due to the disappearance of the $\sqrt{21}$-Ag surface reconstruction and seemingly behave both independently from one atomic terrace to another. 
The origin of the mutual 2DEG and Ag-2DAG confinement %a 2DEG confined on individual terraces and its associated time-related charge/discharge behaviour 
is interpreted as the result of the occurrence of an Ehrlich-Schwoebel diffusion barrier of \unit[155$\pm$10]{meV} below RT and of the interaction between the Ag-2DAG and growing 3D Ag islands.    

\section*{Results}

\begin{figure}
	\centering
	\includegraphics[width=85mm]{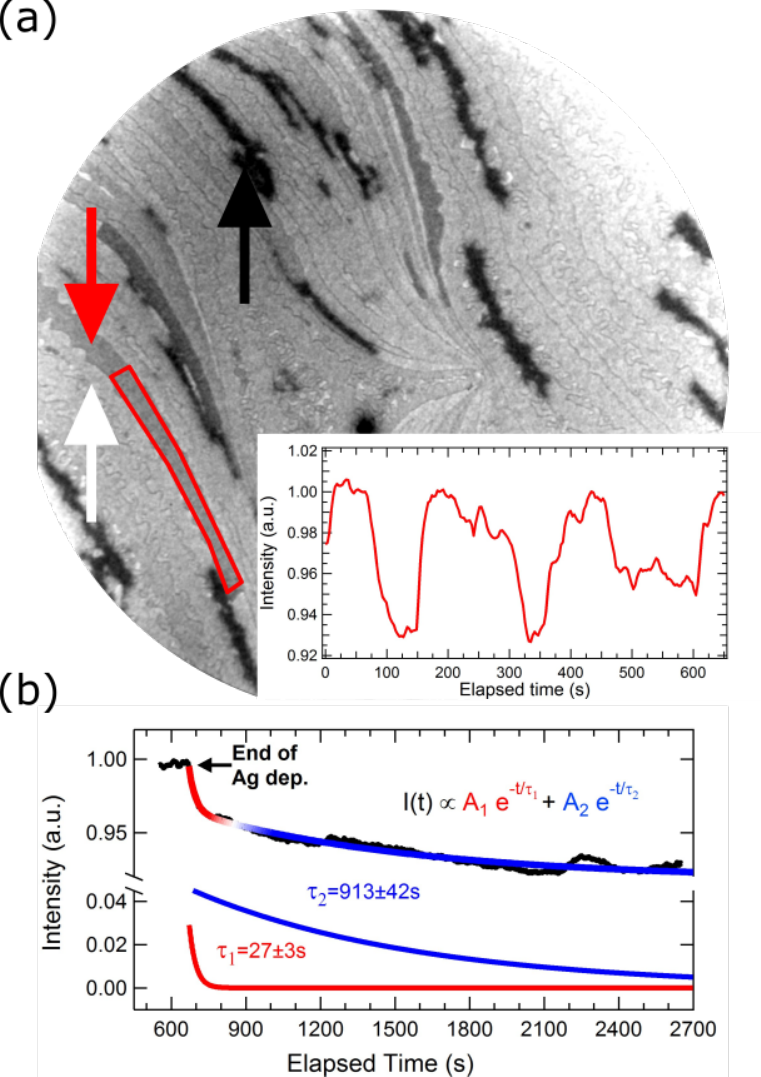}\caption{\label{fig:Inhomo}(a) LEEM image of a $\sqrt{3}$-Ag surface during a Ag deposition at \unit[243]{K} after the nucleation of 3D Ag islands (black areas, see black arrow) showing intensity inhomogeneities as illustrated by the intensity difference between the two neighbouring terraces indicated by the white and red arrows (electron energy, $\varepsilon$=\unit[24]{eV}, field-of-view, FOV=\unit[15]{$\mathrm{\mu}$m}). The inset illustrates the intensity time-evolution of the terrace highlighted in red in (a). 
		(b) Time evolution of the LEEM image intensity averaged over the field-of-view (excluding 3D Ag islands) at the end of a Ag deposition at \unit[220]{K} following the inhomogeneous regime ($\varepsilon$=\unit[24]{eV}). $\tau_1$ and $\tau_2$ are the parameters determined by a double exponential fit of the intensity variation % after the Ag source shutter is closed 
		(red-to-blue curve). The two exponential fits (arbitrary vertical shift) are shown (red and blue curves). % to highlight the different time scales. %The imaging parameters for the inset are: FOV=\unit[25]{$\mu$m} and $\varepsilon$=\unit[24]{eV}.
	}
\end{figure}

In Fig. \ref{fig:Inhomo}, we qualitatively characterise the inhomogeneous regime. Fig. \ref{fig:Inhomo} (a) shows a LEEM image of a $\sqrt{3}$-Ag surface during a Ag deposition at \unit[243]{K}. Apart from growing 3D Ag islands (black areas, see black arrow), the surface intensity is clearly inhomogeneous and varies from one atomic terrace to another (compare for instance the atomic terraces indicated by the red and white arrows, see also Supplementary Video S1 for real-time imaging). The inset evidences large intensity variations observed on an individual terrace as a function of time with quasi-stationary configurations of a few minutes and transitions occurring in only a few tens of seconds. Finally, for samples in the temperature range \unit[210-250]{K}, the reflected intensity averaged over the imaged surface (excluding 3D Ag islands) remains inhomogeneous even in the absence of Ag deposition for times as long as \unit[10-15]{min}. As shown in Fig. \ref{fig:Inhomo}(b), the surface inhomogeneities relax in time unexpectedly with two well-identified exponential timescales, $\tau_1$ and $\tau_2$. The understanding of the inhomogeneous regime (including the two-component relaxation) constitutes the main goal of this report.  %Ag-2DAG concentration decay at the end of the deposition is not a simple exponential time-evolution as expected from nucleation theory\cite{ven84}. For samples deposited above RT, the model using a single characteristic time $\tau$ fits our experiments.

%SM2=2014_11_07_6_LT Dep_15mu_1p8V

\begin{figure}
	\centering
	\includegraphics[width=85mm]{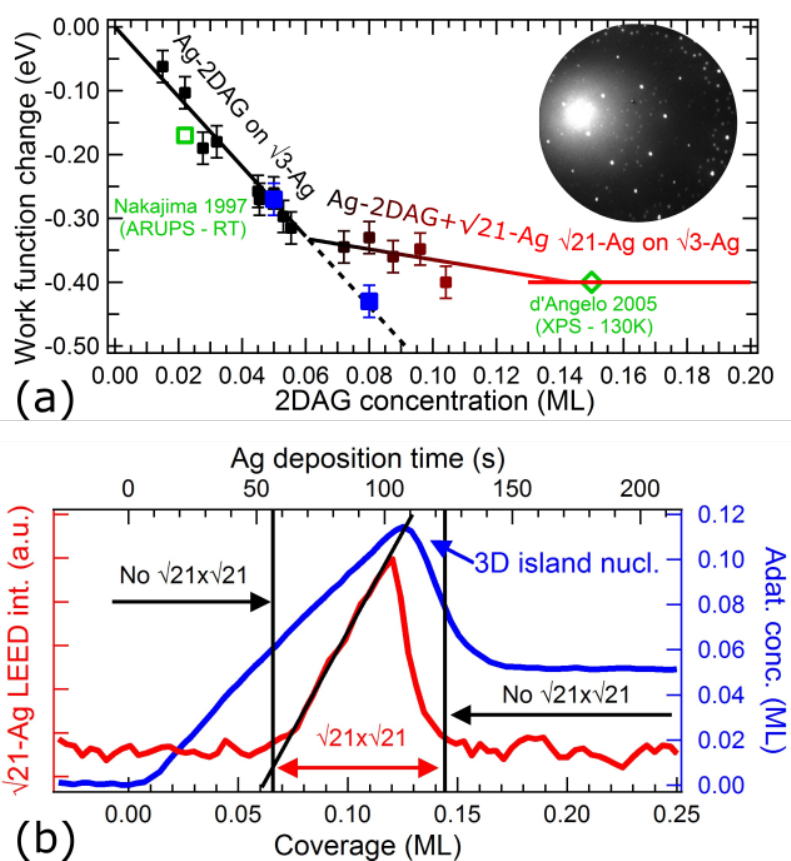}\caption{\label{fig:WF2DAG}
		(a) Surface work function change with respect to the $\sqrt{3}$-Ag surface as a function of the Ag-2DAG concentration % determined by LEEM reflectivity monitoring (black symbols) or using the deposition duration (green symbols) with/without Ag flux (full/open symbols) 
		in the temperature range \unit[210-470]{K}. The lines are data linear fits. Data from the literature are reported in green symbols (Refs. \cite{nak97,dan05}). %The decrease of the $\sqrt{3}$-Ag surface state $S_1$ minimum detected by Y. Nakajima \emph{et al.} under an additional Ag-deposition of \unit[0.022]{ML}\cite{nak97} is reported (red open square) along with the work function diminution found by M. D'angelo \emph{et al.} \cite{dan05} for the $\sqrt{21}$-Ag (red open diamond). 
		The blue symbols %circle (resp.  purple triangle) 
		show the work function change obtained in the inhomogeneous regime %on highly-doped terraces (resp. on the overall surface) in 
		[see the insets in Figs. \ref{fig:WFInhomo}(b, d)]. %The dark red line is a fit reproducing the model. 
		The inset is a $\sqrt{21}$-Ag electron diffraction pattern. (b) Red curve: intensity real-time monitoring of the $\sqrt{21}$-Ag electron diffraction pattern upon Ag deposition at \unit[228]{K} ($\varepsilon$=\unit[16.4]{eV}). Blue curve: real-time monitoring of the Ag-2DAG concentration upon deposition at \unit[210]{K}. The black line is a fit showing the linear dependence of the diffracted intensity of the $\sqrt{21}$-Ag reconstruction with respect to the Ag deposition.}
\end{figure}

Let us first study the effect of the Ag deposition on the  $\sqrt{3}$-Ag surface work function. As described in the Methods, the LEEM technique allows for the determination of the Ag adatom gas (Ag-2DAG) concentration on the $\sqrt{3}$-Ag surface and its induced work function change with respect to the initial $\sqrt{3}$-Ag surface. % In the temperature range investigated here (\unit[210-470]{K}), the behaviour of the Ag-2DAG is not trivial as, apart from a temperature-dependent 3D phase nucleation aforementioned, it forms, below \unit[250]{K} an ordered superstructure identified as a Si(111)$\sqrt{21}\times\sqrt{21}$-Ag \citep{has99}. This superstructure only exists for a finite coverage range (typically \unit[0.1-0.2]{ML} at T$\lesssim\unit[250]{K}$) that widens as the deposition temperature is lowered. Below $\unit[250]{K}$, Ag-2DAG diffusion is sufficiently hindered so that  Si(111)$\sqrt{21}\times\sqrt{21}$-Ag domains nucleate and grow until they reach the corresponding saturation coverage. The Si(111)$\sqrt{21}\times\sqrt{21}$-Ag structure is also characterized by a sharp increase of the surface conductivity at its surface saturation coverage \cite{ton97}. Its band structure differs from that of the Si(111)$\sqrt{3}\times\sqrt{3}$-Ag by a splitting of the surface state $S_\mathrm{1}$ into a dispersive surface state with a large Fermi wave vector thought to be responsible for the large surface conductivity and an adatom-induced localized state\cite{ton01,liu06}. In LEEM, the Si(111)$\sqrt{21}\times\sqrt{21}$-Ag is not directly observed. However its occurrence is clearly detected by LEED. In our low-temperature experiments \unit[210-250]{K}, the Si(111)$\sqrt{21}\times\sqrt{21}$-Ag reconstruction has been observed in the concentration range \unit[0.075-0.14]{ML}. Interestingly, we have clearly established that the Si(111)$\sqrt{21}\times\sqrt{21}$-Ag decay is triggered by the 3D-phase nucleation.  This is most probably caused by the Ag-adatom consumption by the 3D-phase which depletes the 2DAG and destabilizes the Si(111)$\sqrt{21}\times\sqrt{21}$-Ag dense  reconstruction. %\textcolor{red}{Si(111)$\sqrt{21}\times\sqrt{21}$-Ag detected in LEEM at 1.8V ? Origin of reflectivity increase, see \cite{mat05} ??}
%As already reminded, the LEEM technique allows for surface work function determinations using reflectivity $I(\varepsilon)$ measurements. 
%The influence of the Ag-2DAG on the surface work function change with respect to the $\sqrt{3}$-Ag ($\Delta\phi$) is illustrated in Fig. \ref{fig:WF2DAG}(a). 
Fig. S1 illustrates, for instance, that a deposition of \unit[0.04]{ML} of Ag on the $\sqrt{3}$-Ag surface yields a shift of the Intensity-Electron beam energy curve, $I(\varepsilon)$, corresponding to a work function lowering of \unit[-0.23]{eV}. This confirms the electron donor role of the Ag adatoms to the $\sqrt{3}$-Ag surface state reported in Refs. \cite{nak97,uhr02,cra05,liu06}. Data in Fig. \ref{fig:WF2DAG} (a) compile the characterisations of the Ag-2DAG concentration and the surface work function change for various samples in the temperature range \unit[210-470]{K}. %In particular, Ag-2DAG concentrations have been determined by LEEM reflectivity monitoring as previously described (black symbols) or using the deposition duration (green symbols) with/without Ag flux (full/open symbols). Apart  all LEEM experiments clearly fall into a universal curve which demonstrates the generality of the results established here. 

For concentrations below \unit[0.06]{ML}, the work function decreases linearly as the Ag-2DAG concentration increases, with a slope of $\Delta\phi_{\sqrt{3}}=$\unit[-5.54$\pm$0.35]{eV/ML} (black line). For comparison, we add in Fig. \ref{fig:WF2DAG} the diminution of the $S_1$ surface state minimum of \unit[0.17]{eV} measured by Y. Nakajima \emph{et al.} in Ref. \cite{nak97} using Angle-Resolved UV-Photoemission Spectroscopy after a Ag-additional deposition of \unit[0.022]{ML}. This value is in quantitative agreement with our data and confirms that the work function measurements characterise the 2DEG filling. 

%2014-05-21
%WF Changes
%2014-09-17/2014-09-02

Above \unit[0.06]{ML}, we evidence a breakdown of the initial linear relationship between the measured quantities [see the slope change between the black and the black-to-red lines in Fig. \ref{fig:WF2DAG}(a)]. %Also we have measured a work function lowering of \unit[0.4]{eV} for a Ag-2DAG concentration of \unit[0.11]{ML}. 
These Ag-2DAG concentrations can only be obtained below RT (\unit[210-250]{K}). In this temperature range, a superstructure, identified by electron diffraction as $\sqrt{21}$-Ag \cite{zha95}, appears % for Ag-2DAG concentrations \unit[$>0.06$]{ML}
 [see the inset of Fig. \ref{fig:WF2DAG}(a)]. This surface reconstruction is characteristic of a periodic arrangement of the Ag-2DAG on an preserved $\sqrt{3}$-Ag surface that only exists for a temperature-dependent finite coverage range \cite{has99,fuk06}. The complete coverage of the surface by the $\sqrt{21}$-Ag reconstruction is expected at \unit[0.143]{ML} (\emph{i.e.} $1/7$ of a ML and three Ag atoms per $\sqrt{21}\times\sqrt{21}$-unit cell). 
 
 To confirm the role of the $\sqrt{21}$-Ag reconstruction in the observed regime change, we have measured in real time the $\sqrt{21}$-Ag electron diffraction pattern (Fig. \ref{fig:WF2DAG}(b), red curve). We compare, under similar deposition temperature, its time-dependent intensity to a Ag-2DAG concentration monitoring obtained from the real-space imaging of the surface (Fig. \ref{fig:WF2DAG}(b), blue curve). We clearly evidence: \emph{(i)} the $\sqrt{21}$-Ag electron diffraction pattern appears above \unit[0.06]{ML}, \emph{(ii)} its intensity increases linearly with the Ag deposition (black line) and \emph{(iii)} its decay is triggered by the temperature-dependent 3D Ag island nucleation. From this, we can draw that the $\sqrt{21}$-Ag surface reconstruction requires a critical supersaturation before the nucleation of  $\sqrt{21}$-Ag domains. The $\sqrt{21}$-Ag surface reconstruction coverage extends until 3D Ag islands nucleate. The maximum of the $\sqrt{21}$-Ag reconstruction coverage is determined by the temperature-dependent 3D Ag island nucleation. The complete coverage of the surface by the $\sqrt{21}$-Ag reconstruction at \unit[0.143]{ML} needs deposition temperatures below \unit[210]{K} to be reached. The observed decay of the  $\sqrt{21}$-Ag reconstruction is caused by the growing 3D islands which consumes adatoms and destabilizes the $\sqrt{21}$-Ag dense reconstruction. Finally, our high concentration measurements [$\sim\unit[0.1]{ML}$, Fig. \ref{fig:WF2DAG}(a)] are compatible with a work function change of $\Delta\phi_{\sqrt{21}}=$\unit[-0.4]{eV} obtained by D'angelo \emph{et al.} in Ref. \cite{dan05} using X-ray Photoemission Spectroscopy measurements at \unit[130]{K} for a fully-covering $\sqrt{21}$-Ag reconstruction% and exhibiting a $\sqrt{21}$-Ag diffraction pattern
. The deviation from the initial linear dependence of the charge transfer with respect to the Ag-2DAG concentration and the saturation of the work function change at \unit[-0.4]{eV} are thus clearly attributed to the occurrence of the $\sqrt{21}$-Ag surface reconstruction. % with an increasing partial coverage of the $\sqrt{21}$-Ag over the $\sqrt{3}$-Ag. The $\sqrt{21}$-Ag surface reconstruction completely covers the surface  % in the concentration range \unit[0.06-0.14]{ML}. 
This is most probably due to the electronic localised state % in the vicinity of the positively-charged Ag adatoms 
observed below RT in Refs. \cite{cra05,liu06} which is possibly at the origin of the localised $\sqrt{21}$-Ag $D$ state found in Ref. \cite{ton01}% and/or the band folding of the $\sqrt{21}$-Ag Fermi surface evidenced in Ref. \cite{mat05})
. 

\begin{figure}[h!]
	\centering
	\includegraphics[width=80mm]{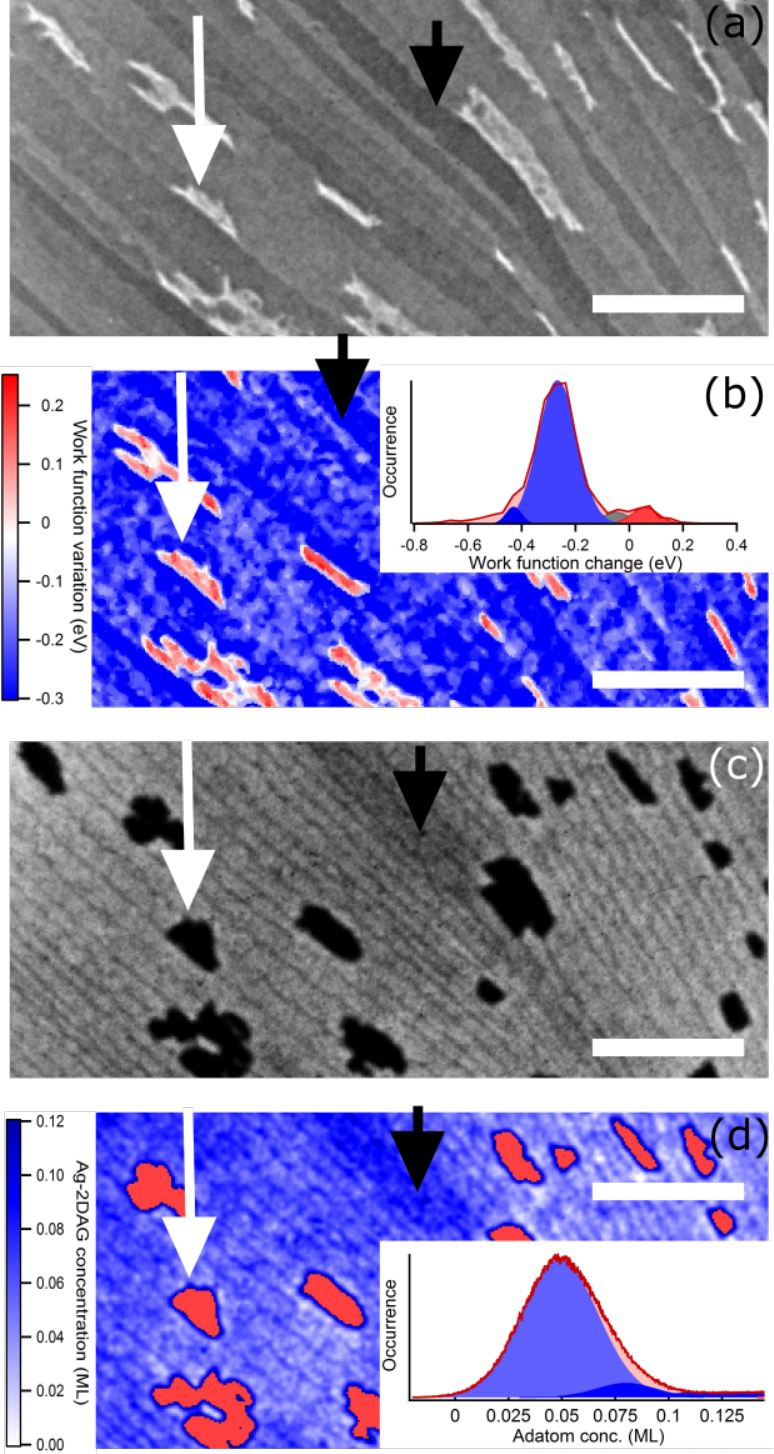}\caption{\label{fig:WFInhomo} (a) Mirror Electron Microscopy image of a $\sqrt{3}$-Ag surface during a 2DAG-Ag deposition at \unit[220]{K} after the nucleation of 3D island (white intensity) showing intensity inhomogeneities ($\varepsilon$=\unit[0]{eV}). (b) 2D map of the surface work function change with respect to the $\sqrt{3}$-Ag surface obtained by reflectivity measurements (see Methods) % as a function of the incident electron beam energy
		. The inset shows the histogram of the work function map. (c) LEEM image ($\varepsilon$=\unit[24]{eV}) acquired \unit[8]{min} before Fig. \ref{fig:WFInhomo}(a-b). (d) The 2D map of the Ag-2DAG concentration derived from image (c) (see Methods). The 3D Ag islands have been coloured in red in Fig. \ref{fig:WFInhomo}(d). The inset shows the histogram of the Ag adatom gas concentration. See the text for the arrow meaning. For all images, the scale bar is \unit[2]{$\mu$m}.}
\end{figure}

To understand the deviation from the initial $\Delta\phi(c)$-linear relationship induced by the occurrence of the $\sqrt{21}$-Ag surface reconstruction, we propose a simple interpretation based on the mean work function change with respect to the initial $\sqrt{3}$-Ag surface. The work function change, $\Delta\phi$, is derived as the average between the contributions of both surface reconstructions (defined as $\Delta\phi_{\sqrt{3}}(c)$ and $\Delta\phi_{\sqrt{21}}=\unit[-0.4]{eV}$ \cite{dan05}) weighted by their respective area fraction. We assume that above the $\sqrt{21}$-Ag nucleation concentration, $c_{\mathrm{n,\sqrt{21}}}$, additional Ag adatoms only contribute to the growth of the $\sqrt{21}$-Ag domains. In other words, a Ag-2DAG of concentration $c_{\mathrm{n,\sqrt{21}}}$ coexist with a growing $\sqrt{21}$-Ag surface reconstruction. The $\sqrt{21}$-Ag reconstruction is indeed interpreted as a condensed configuration of the Ag-2DAG above a preserved $\sqrt{3}$-Ag surface \cite{fuk06}. In this model, $\Delta\phi$ reads: 

%\begin{equation*}
%\Delta\phi=\Delta\phi_{\sqrt{21}}\;\rho_{\sqrt{21}} + \Delta\phi_{\sqrt{3}}(c_{\mathrm{n,\sqrt{21}}})\;(1-\rho_{\sqrt{21}})
%\end{equation*}

where $\rho_{\sqrt{21}}=(c-c_{\mathrm{n,\sqrt{21}}})/(c_{\mathrm{m,\sqrt{21}}}-c_{\mathrm{n,\sqrt{21}}})$ is the area fraction covered by the $\sqrt{21}$-Ag reconstruction and $c_{\mathrm{m,\sqrt{21}}}=1/7=\unit[0.143]{ML}$ \cite{fuk06,mat07b} is the Ag-2DAG concentration for which the $\sqrt{21}$-Ag reconstruction reaches the maximum fraction area (\emph{i.e.} $\rho_{\sqrt{21}}(c_{\mathrm{m,\sqrt{21}}})=1$). The linear model is shown as a black-to-red line in Fig. \ref{fig:WF2DAG}(a) and fits the experimental data for $c_{\mathrm{n,\sqrt{21}}}=\unit[0.061\pm0.005]{ML}$. This threshold value for the onset of the $\sqrt{21}$-Ag surface reconstruction is in quantitative agreement with the occurrence of a $\sqrt{21}\times\sqrt{21}$ electron diffraction pattern at a value of $c\simeq\unit[0.06]{ML}$ [Fig. \ref{fig:WF2DAG}(b)] which confirms the validity of our approach. %It also reproduces the change of slope shown in Fig. \ref{fig:WF2DAG}(b) and thus explains the breakdown of the linear relationship between the Ag-2DAG concentration and the surface work function.  %This concentration value is lower than the threshold of \unit[0.072]{ML} for the occurrence of the $\sqrt{21}$-Ag diffraction pattern. This means that the $\sqrt{21}$-Ag is detected only above an area coverage of $\rho_{\sqrt{21}}\simeq0.13$. This value is higher than the diffraction technique detection limit ($\lesssim0.1$). This suggests that a supersaturation of $k_{\mathrm{B}}\mathrm{T}\ln(0.072/0.061)\sim$ \unit[3]{meV} to nucleate the $\sqrt{21}$-Ag reconstruction on the $\sqrt{3}$-Ag surface is required in this temperature range. 
The onset of the inhomogeneous regime observed upon Ag deposition below RT follows the nucleation of the 3D Ag islands and the disappearance of the $\sqrt{21}$-Ag. Fig. \ref{fig:WFInhomo}(a) and Supplementary Information show that sub-micron wide terraces can exhibit an intensity darker  %or brighter (white arrow) 
than the surface mean intensity over micron-scaled lengths (see black arrow) with time fluctuations. In-between transitions, the temporal evolution of the surface is sufficiently slow to allow for the determinations of the 2D work function map [Fig. \ref{fig:WFInhomo}(b)] and the Ag 2D concentration map [Fig. \ref{fig:WFInhomo}(d)]. % shows the spatial work function deviations from the Si(111)$\sqrt{3}\times\sqrt{3}$-Ag reference ($\Delta\phi$). 
Three distinct features can be identified in the 2D work function map and its histogram [see the inset of  Fig. \ref{fig:WFInhomo}(b)]. Most of the surface exhibits a work function change with respect to a homogeneous $\sqrt{3}$-Ag surface of $\Delta\phi\simeq$\unit[-0.27]{eV} (light blue). 3D islands have a work function higher than the reference by typically \unit[0.1-0.2]{eV} (white arrow). A $\sqrt{3}$-Ag work function value close to \unit[4.55]{eV} \cite{dan05,hug05} gives an absolute work function for the 3D islands in quantitative agreement with the value of \unit[4.64]{eV} (\unit[4.72]{eV} resp.) reported for bulk Ag(100) (Ag(111) resp.) using photoelectric measurements \cite{dwe75}. %This confirms \emph{a posteriori} the occurrence of these crystal orientations in the 3D-Ag island colonies\cite{abu95}. 
Large %(resp. narrow) 
terraces with a dark blue %(resp. white) 
intensity (black arrow) are also observed and are characterized by $\Delta\phi\simeq$\unit[-0.43]{eV}% (resp. $\Delta\phi\simeq$\unit[-0.05]{eV})
. 

%2015-01-26_WF 2Dmap

These % last three
features % are in direct connection with the Ag-2DAG and 
can be better understood by looking at the Ag adatom gas concentration map % component of the system
%. The 2D concentration map of the Ag-2DAG obtained by a concentration monitoring upon deposition is
and its histogram [see the inset of  Fig. \ref{fig:WFInhomo}(d)]. The Ag-2DAG concentration and the 2D work function maps have been acquired with a time delay [typ. \unit[8]{min} between Figs. \ref{fig:WFInhomo}(a-b) and Figs. \ref{fig:WFInhomo}(c-d)]. This results in slightly different spatial and size distributions of the Ag 3D islands. Also the time delay makes the one-to-one correspondence between the terrace work function changes and the local Ag-2DAG concentration difficult. However both 2D maps and their respective histograms exhibit convincing similarities% active areas
. In particular, dark blue terraces observed in Fig. \ref{fig:WFInhomo}(b) and characterised by $\Delta\phi\simeq\unit[-0.43]{eV}$ appear also as dark blue terraces in Figs. \ref{fig:WFInhomo}(c-d) with a Ag adatom concentration $c\simeq\unit[0.08]{ML}$, as indicated by the black arrow. With the disappearance of the $\sqrt{21}$-Ag surface reconstruction, this data point ($c\simeq\unit[0.08]{ML}$, $\Delta\phi\simeq\unit[-0.43]{eV}$) is unexpectedly in agreement with the initial $\Delta\phi(c)$-linear relationship found for homogeneous surfaces [Fig. \ref{fig:WF2DAG}(b)]. When averaging over multiple atomic terraces, we derive a mean surface Ag-2DAG concentration of \unit[$c\simeq$0.05]{ML} yielding a mean work function change of $\Delta\phi\simeq\unit[-0.27]{eV}$ which also perfectly matches the initial $\Delta\phi(c)$-linear relationship. Data obtained in the inhomogeneous regime are reported in Fig. \ref{fig:WF2DAG} as blue symbols. This demonstrates two results: (i) the observed temporal and spatial changes of the surface work function mirror the variations of the Ag-2DAG concentration (\emph{i.e.} atomic terraces highly concentrated in Ag adatoms exhibit a high charge transfer). 
(ii) In the inhomogeneous regime, work function changes lower than \unit[-0.4]{eV} can be reached owing to the disappearance of the $\sqrt{21}$-Ag. The initial linear work function change/Ag adatom concentration relationship obtained on homogeneous surfaces is still locally verified in the inhomogeneous regime. %the linear work function change/concentration relationship obtained on homogeneous surfaces (Fig. \ref{fig:WF2DAG}) remains quantitatively valid in the inhomogeneous regime even for concentration values where the onset of the Si(111)$\sqrt{21}\times\sqrt{21}$-Ag reconstruction should limit the work function change.  %\textcolor{red}{$\Delta\phi_{\mathrm{mean}}$\unit[$\simeq$-0.27]{eV}  vs $c_{\mathrm{mean}}$ }. %The main component of the 2D map comes from the stationary Ag-2DAG concentration of \unit[$\simeq$0.05]{ML} (equivalently $\Delta\phi$\unit[$\simeq$-0.27]{eV}). Large (resp. narrow) terraces with $\Delta\phi\simeq\unit[-0.43]{eV}$ (resp. $\Delta\phi\simeq\unit[-0.06]{eV}$) are actually areas of Ag accumulation (resp. Ag depletion). The narrow depletion areas are mainly observed between two growing island colonies which consume Ag adatoms. Terraces with Ag-2DAg accumulation are generally located close to depletion areas to provide Ag adatoms. However mass transfer seems to be hindered probably by a step-induced Ehrlich-Schwoebel barrier. This clearly demonstrate the inhomogeneity of the Ag-2DAG at the mesoscopic scale.  

\section*{Discussion}

In the following, we provide an explanation of the inhomogeneous regime based on experimental measurements that rely on the dependence of % the electronic properties of 
the charge transfer on the underlying Ag atomic processes (adsorption, diffusion, capture). In this regime, the Ag-2DAG concentration (resp. surface work function change) can locally increase (resp. decrease) by \unit[60]{\%} in only \unit[$\sim$30]{s} on atomic terraces as large as \unit[$\gtrsim$3]{$\mu$m$^2$}. %As already said, when the incoming atomic flux is stopped, the Ag-2DAG concentration decreases until it reaches the value describing the thermodynamical equilibrium between the diffusing 2DAG and the growing 3D phase. 
Also it is worth highlighting that the inhomogeneous regime, observed upon Ag deposition below RT, follows the 3D island nucleation. This suggests that a non-equilibrium mass transfer between the growing 3D islands and the surrounding Ag-2DAG phase is at work.

 In the data obtained at \unit[220]{K} [Fig. \ref{fig:Inhomo}(b)], the Ag-2DAG concentration relaxation after the end of the Ag deposition ($t=\unit[670]{s}$) is not a simple exponential time-evolution as expected from nucleation theory\cite{ven84}. Indeed, when adatoms are only captured by stable growing 3D islands, $\partial n_1/\partial t=-\sigma_xDn_xn_1$, where $n_1, \sigma_x, D$ and $n_x$ are respectively the adatom surface concentration, the adatom capture number, the adatom diffusion coefficient and the stable-island density. This gives $n_1(t)\sim e^{-t/\tau}$ with $\tau=1/\sigma_x D n_x$ being the adatom relaxation time. In the inhomogeneous regime, two exponential functions with different timescales are required to obtain satisfying fits to the experimental data. %The physical origin of these short/long time-scale phenomena are now to be discussed.
 This implies that additional activation barriers are involved below RT. One may think of the Ehrlich-Schwoebel barrier to cross atomic steps that would confine adatoms on terraces% and/or interactions between the charged Ag adatoms
. %To quantify this additional activation barrier, 

\begin{figure}[h!]
	\centering
	\includegraphics[width=85mm]{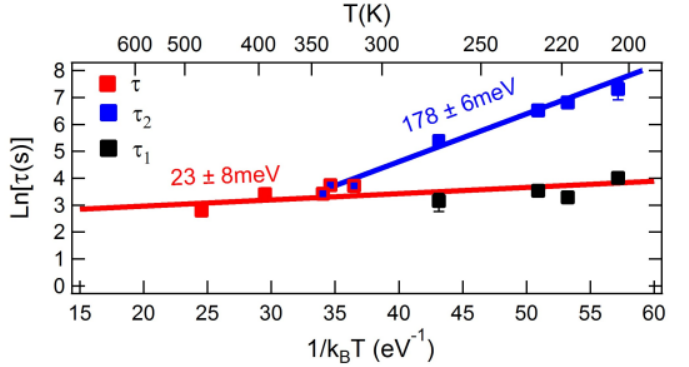}\caption{\label{fig:Tau}  
		Arrhenius plot of the Ag-2DAG concentration relaxation time ($\tau$) in the temperature range \unit[210-470]{K}. Data obtained below RT require two characteristic times, $\tau_1$ and $\tau_2$, to fit satisfyingly the experimental data (see text for details).}
\end{figure}

%AdatConc	

To confirm this hypothesis, we have characterised the Ag-2DAG concentration decay in the temperature range \unit[210-470]{K}. In Fig. \ref{fig:Tau}, we display the single characteristic time, $\tau$, obtained for depositions above RT and the two timescales, $\tau_1$  and $\tau_2$, determined below RT as a function of the deposition temperature. % Fig. \ref{fig:Tau} reveals that below RT, $\tau_1$  and $\tau_2$ have different behaviours. 
The parameter $\tau_1$ appears as the continuation of the characteristic time $\tau$ that characterises the capture of Ag adatoms by Ag 3D islands in the homogeneous regime. As such the short-timescale evolution of the Ag-2DAG concentration exhibits a relatively weak activation energy of \unit[23$\pm$8]{meV} over the explored temperature range (Fig. \ref{fig:Tau}, red linear fit). %is negligibly dependent on the temperature ($\tau_1\sim\unit[30]{s}$) while $\tau_2$ exhibits an activation energy of \unit[178$\pm$5]{meV}. 
On the other hand, $\tau_2$ shows a more significant dependence on the temperature with an activation energy of \unit[178$\pm$6]{meV} (Fig. \ref{fig:Tau}, blue linear fit). As $\tau_2$ governs the relaxation dynamics below RT, the activation energy of the concentration relaxation time is increased by \unit[155$\pm$10]{meV} below RT. This increase compares remarkably well to the Ehrlich-Schwoebel barrier of Ag/Ag(111) self-diffusion of \unit[150$\pm$30]{meV} \cite{vri94}, \unit[120$\pm$20]{meV} \cite{bro95} and \unit[130$\pm$40]{eV} \cite{mor98} determined below or close to RT. In the low-temperature regime, high Ag adatom incorporation by attachment at atomic steps is likely due to a reduced surface diffusion. We can thus expect in this temperature regime a behaviour similar to the Ag/Ag(111) system. At higher temperatures, lateral strain relaxations at atomic steps are possible. This favours atom exchange mechanisms at steps that have been shown to be the preferred route over jump diffusion in the Ag/Ag(111) system \cite{fer97} and in a very similar system Au/Si(111)$\sqrt{3}\times\sqrt{3}$-Au \cite{art-CHE16c}. %simple temperature-based arguments explain the occurrence of an additional Ehrlich-Schwoebel barrier below RT. In the low-temperature regime, high Ag adatom incorporation at atomic steps is likely due to a reduced surface diffusion. %The activation energy increase determined in this study compares remarkably well to values reported in the literature about the Ehrlich-Schwoebel barrier of Ag/Ag(111) self-diffusion of \unit[0.12$\pm$0.02]{eV} \cite{bro95} and \unit[0.13$\pm$0.04]{eV} \cite{mor98} determined in similar temperature ranges. 
Although charge-dependent diffusion of Ag adatoms cannot be excluded \cite{rep16}, the above arguments strongly suggest that below RT, an additional Ehrlich-Schwoebel diffusion barrier limits the Ag-2DAG concentration equating between terraces as well as the induced charge transfer and explains the long timescale behaviour of these quantities and specified by $\tau_2$. This results in the confinement of adatoms on atomic terraces. 
%Neighbouring terraces behave independently unless a 3D island grows at the boundary between two terraces. 

\begin{figure}[h!]
	\centering
	\includegraphics[width=70mm]{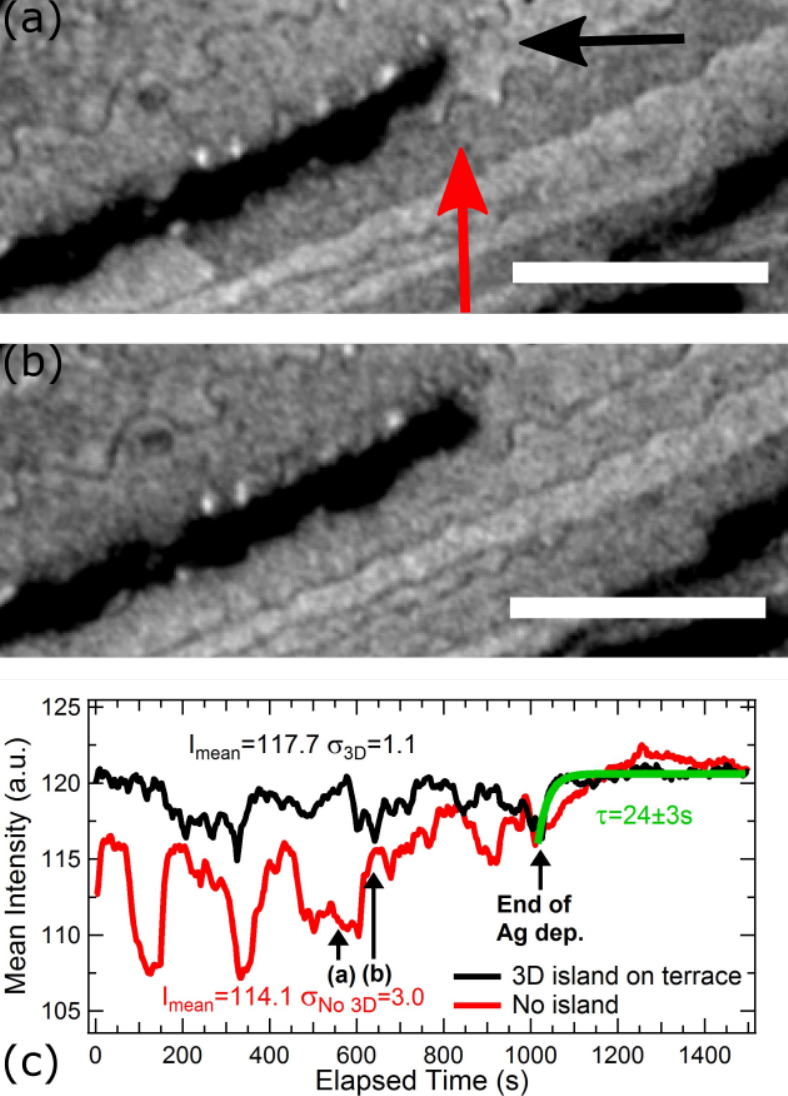}\caption{\label{fig:TimeEvol} (a-b) Snapshots [from Fig. \ref{fig:Inhomo}(a)] of the growth of a 3D island (black intensity) before (a) and after (b) it crosses the delimiting step with the terrace marked by the red arrow. Between (a) and (b), the adatom concentration of the terrace marked by the red arrow has decreased toward its stationary value ($c_{\mathrm{stat}}\simeq\unit[0.04]{ML}$). The scale bar is \unit[1.5]{$\mu$m} and $\varepsilon$=\unit[24]{eV}. (c) Intensity time-evolution of the atomic terraces indicated in (a) by red and black arrows. The timeline of the snapshots (a) and (b) is displayed.}
\end{figure}

Besides the Ehrlich-Schwoebel diffusion barrier, the growth of 3D islands plays a major role in the occurrence of the inhomogeneous regime. Figs. \ref{fig:TimeEvol} (a-b) are snapshots of Fig. \ref{fig:Inhomo}(a) showing the growth of a 3D island (black intensity) on a terrace (black arrow) before and after it crosses a delimiting step and reaches a supersaturated neighbouring terrace (red arrow). The intensity of the terrace indicated by the red arrow clearly changes from dark grey to light grey suggesting a decrease of the Ag-2DAG concentration. This change is visible in the intensity time evolution of the terrace [Fig. \ref{fig:TimeEvol} (c)] at \unit[$t\simeq$600]{s}. The Ag-2DAG accumulated on the terrace indicated by the red arrow relaxes towards its stationary concentration ($c_{\mathrm{stat}}\simeq\unit[0.04]{ML}$) when the 3D island crosses the delimiting atomic step. Also, under Ag atom deposition, the intensity variation of a terrace occupied by a growing 3D island (Fig. \ref{fig:TimeEvol} (c), black curve) is much smaller than that of an unoccupied terrace (Fig. \ref{fig:TimeEvol} (c), red curve)% as indicated by the intensity standard deviation $\sigma$ (\emph{i.e.} $\sigma_{\mathrm{3D}}<\sigma_{\mathrm{No\ 3D}}$)
. High Ag-2DAG concentrations and charge transfer on atomic terraces exhibiting a 3D island cannot be reached owing to the adatom consumption by the growth of the 3D islands. Finally, atomic terraces hosting a growing 3D island show a rapid Ag-2DAG/charge transfer relaxation of \unit[$\sim$30]{s} when the Ag flux is stopped (see for instance the black curve in Fig. \ref{fig:TimeEvol}(c) with a characteristic time of \unit[24$\pm$3]{s}) that compares well with the characteristic time $\tau_1$. From these local measurements, we can infer that the short-time behaviour observed in the inhomogeneous regime and quantified by $\tau_1$ corresponds to the capture of diffusing Ag adatoms by Ag 3D islands in their vicinity. This also confirms \emph{a posteriori} that $\tau$ and $\tau_1$ are the same quantity and clock the same phenomenon. % In the explored temperature range $\tau_1$ appears as temperature independent. This suggests that the adatom capture by the 3D islands in their vicinity may be limited by the average terrace size. % The relaxation of the Ag-2DAG concentration/2DEG charge transfer induced by the consumption of a supersaturated Ag-2DAG by 3D islands to reach the 3D island/Ag-2DAG thermodynamic equilibrium. 
The occurrence, below RT, of two distinct timescales ($\tau_1$ and $\tau_2$) resulting from the interplay between a Ag adatom gas with a hindered surface diffusion and a growing 3D phase convincingly explains the large variations of the Ag-2DAG concentration and 2DEG charge transfer.% detected through the surface work function. 

%2014-05-21_movie5_bis	

\section*{Conclusion}
	In summary, we have monitored, upon Ag-deposition on a $\sqrt{3}$-Ag, the Ag-2DAG concentration and the corresponding surface work function change using LEEM. We have demonstrated that both quantities are linearly related and that the breakdown of the linear relationship is caused by the onset of the $\sqrt{21}$-Ag below RT. An unconventional regime showing inhomogeneous spatial distribution and temporal variations of both work function change and Ag-2DAG concentration is evidenced. The interaction between the growing 3D Ag islands and the Ag-2DAG plays a major role in the origin of this regime along with the occurrence of an Ehrlich-Schwoebel barrier of \unit[155$\pm$10]{meV} below RT. Inhomogeneous 2D electron gas in a quasi-permanent time configuration are likely to be obtained by quenching the system under study. These results are of high interest for the emerging field of low dimensional electronics.

\section*{Methods}
\label{sec:methods}

\subsection*{Sample preparation}
%\label{subsec:sample}
\emph{n}- and \emph{p}-doped Si(111) samples %with nominal resistivity \unit[1-10]{$\Omega$cm} 
have been cleaned %with acetone and ethanol using ultrasonic bath and dry cleaned with nitrogen 
before loading into our ultra-high vacuum setup \cite{art-CHE13g} for a complete characterization by Low-Energy Electron Microscopy (\emph{LEEM III}, Elmitec GmbH). The sample temperature is determined using a type-C thermocouple spot-welded to the sample holder. We estimate that the typical measurement uncertainty is $\pm\unit[25]{K}$. %The use of both \emph{n}- and \emph{p}-doped Si(111) samples makes sure that the doping of the substrates is irrelevant in the current study. %Samples have been degassed for \unit[1-2]{h} at \unit[$\sim$850]{K} in a dedicated chamber. ,
After a degas annealing (\unit[1-2]{h} at \unit[$\sim$850]{K}), samples have been flash heated at \unit[1400-1500]{K} for a few tens of seconds and produced high quality $7\times7$ surface reconstruction (not shown). %Prior to Ag deposition in the LEEM chamber. 
%This produced Si(111) surfaces of high quality exhibiting a $7\times7$ surface reconstruction (not shown). The surface used as a substrate for the 2DEG characterization is a  %Ag-2D adatom gas (2DAG) is a 
$\sqrt{3}$-Ag reconstructed surfaces are obtained by Ag deposition on a Si(111) surface below \unit[750]{K} %. %Ag has been deposited using a Knudsen-effusion cell including an alumina crucible that contained 5N Ag shots. 
%Ag deposition rate has been calibrated using the time required to obtain a complete $\sqrt{3}$-Ag surface as observed in LEEM 
(typically \unit[15]{min} for \unit[1]{ML}-$\sqrt{3}$-Ag with \unit[1]{ML}=\unit[7.83$\times$10$^{14}$]{atoms/cm$^{2}$} \cite{nak97}). %This monolayer deposition is fulfilled for an atom density equal to that of a Si(111) surface, \emph{i.e.} \unit[1]{ML}=\unit[7.83$\times$10$^{14}$]{cm$^{-2}$} \cite{nak97}. 
%Typical deposition rates of \unit[1]{ML}-Ag in \unit[15]{min} have been used (\unit[1]{ML}=\unit[7.83$\times$10$^{14}$]{atoms/cm$^{2}$} \cite{nak97}).  %\unit[0.067]{ML/min} have been used (\unit[1]{ML}-Ag in \unit[15]{min}). 
2DEG \emph{per se} is obtained by an additional Ag deposition in the range \unit[210-470]{K} under identical conditions.

\subsection*{LEEM \& work function measurements}
%\label{subsec:methods}
Low-Energy Electron Microscopy (LEEM) technique allows %it is possible 
to determine the work function change of a surface (\emph{e.g.} under deposition conditions) using Intensity-Electron beam energy, $I(\varepsilon)$, curves obtained from reflectivity measurements as a function of the incident electron beam energy averaged over areas exhibiting identical intensity %(typically \unit[1-10]{$\mu$m$^2$})
\cite{bab80,mur12,jin14}. %The work function of the clean substrate can be used as a reference to account for the work function of the electron gun. In the case of the Si(111)$\sqrt{3}\times\sqrt{3}$-Ag surface, previous studies by X-ray Photoelectron Spectroscopy have reported a work function value of \unit[4.55]{eV} \cite{dan05,hug05} which is very close to the value of the Si(111) $7\times7$ surface of \unit[4.6]{eV} \cite{mon01}. %Beside the sophiscated method described in Ref. \cite{bab80} and based on a deconvolution approach, 
%The sample work functions have been determined from the incident electron energy at the intersection of the two linear fits of the reflectivity curve obtained below and above the total reflectivity regime [Fig. \ref{fig:methods} (a)]. In our case, this threshold also matches the condition $I/I_\mathrm{0}<0.92$, where $I_\mathrm{0}$ is the intensity backscattered from the surface in the total reflection regime.     
%The current report completes an approach initiated in Ref. \cite{rau12} and recently amended in Ref. \cite{kau15} to determine the work function 2D map of a surface based on LEEM $I(\varepsilon)$ measurements. 

At low incident electron beam energy, the sample work function is derived from the intersection of the two linear fits of the reflectivity curve obtained below and above the total reflection threshold. With this method, we estimate that the typical measurement uncertainty is \unit[$\pm$25]{meV}. In Fig. S1, a $\sqrt{3}\times\sqrt{3}$-Ag surface shows at \unit[220]{K} an electron beam injection threshold slightly above $\varepsilon=\unit[0]{eV}$. Upon a Ag-2DAG deposition of \unit[0.04]{ML} at \unit[220]{K} a lowering of the electron beam injection threshold of $\Delta\phi=\unit[-0.23]{eV}$ with respect to the $\sqrt{3}\times\sqrt{3}$-Ag surface is evidenced.

%\begin{figure}[h!]
%	\centering
%	\includegraphics[width=90mm]{Fig8a}\caption{\label{fig:methodsWF} Intensity-Incident electron beam energy $I(\varepsilon)$-reflectivity curve of a $\sqrt{3}\times\sqrt{3}$-Ag surface at \unit[220]{K} (black curve). Upon a \unit[0.04]{ML}-Ag deposition at \unit[220]{K} (red curve), the  $I(\varepsilon)$-curve shows a lowering of the electron injection threshold of $\Delta\phi=\unit[-0.23]{eV}$. 
%	}
%	%2015-01-16
%	%Inset 2014-05-14
%\end{figure}

%In our case, 
In both cases, the injection threshold also matches the condition $I/I_\mathrm{0}<0.92$, where $I_\mathrm{0}$ is the intensity backscattered from the surface in the total reflection regime. To determine unambiguously a surface work function using LEEM, a reference is needed to account for the work function of the electron gun. In the case of the Si(111)$\sqrt{3}\times\sqrt{3}$-Ag surface, previous studies by X-ray Photoelectron Spectroscopy have reported a work function value of \unit[4.55]{eV} \cite{dan05,hug05} which is very close to the value of the Si(111) $7\times7$ surface of \unit[4.6]{eV} \cite{mon01}. %Beside the sophiscated method described in Ref. \cite{bab80} and based on a deconvolution approach, 

Instead of averaging the reflectivity curves over micron-sized areas% as previously described
, the work function can be also derived pixel-by-pixel for a given $I(\varepsilon)$-image stack % obtained for different incident electron energy
%. Work function 2D maps are computed 
using the condition $I/I_\mathrm{0}<0.92$ as a criterion in the algorithm. To enhance the signal-to-noise ratio, a $3\times3$ average filter is employed. This procedure provides work function 2D maps \cite{rau12,kau15} % For the work function 2D map, the condition $I/I_\mathrm{0}<0.92$ is used as a criterion in the algorithm. To enhance the signal-to-noise ratio, a $3\times3$ average filter has been employed (see Fig. \ref{fig:WF2DAG}). 
%This method 
and paves the way for in-lab studies of surfaces showing inhomogeneously distributed work function. % without the need of synchrotron radiation-based measurements and/or an energy-filtered detector as in the case of X-ray PhotoEmision Electron Microscopy \cite{rau12}. %in the same manner as Kelvin probe measurements with Kelvin probe force microscopy (KPFM).
%\begin{figure}[h]
%\includegraphics[width=80mm]{Figs/fig1}\caption{\label{fig:methods} (a) Reflectivity (\emph{I-V}) curve of a Ag-2DAG deposited at \unit[220]{K} showing the electron injection threshold at $E=\unit[-0.3]{eV}$. %2015-01-26_WF 2Dmap
% (b) Surface reflectivity change induced by a Ag-2DAG deposition at RT and the deduced Ag-2DAG concentration monitoring (see text for details). The dark line is a linear fit used to derive the value of $\Sigma$ \cite{far98}. The inset shows the reflectivity curves of a $\sqrt{3}\times\sqrt{3}$-Ag surface before and after a \unit[0.09]{ML} Ag-deposition at \unit[220]{K}. A clear sensitivity to the deposition is observed at \unit[24]{eV}. }
%%2015-01-16
%%Inset 2014-05-14
%\end{figure}

\subsection*{LEEM \& adatom concentration monitoring}

%Another significant breakthrough of the LEEM technique is based on approaches developed in the context of the backscattering of atomic beams \cite{far98}. 
Another major asset of the LEEM technique is that it enables to monitor the concentration of an adspecies in real-time\cite{fig06,log09,sch12}. %By comparing the LEEM $I(\varepsilon)$ curves of a sample before and after the deposition, a maximum of sensitivity to the deposited species can be determined. 
The adspecies concentration, $c$, % (in fraction of a ML)
is obtained from the real-time monitoring of the reflectivity changes of a sample region as a function of the deposit, $\theta$, 
at an adequate incident electron energy. This method is based on an approach developed for backscattered atomic beams \cite{far98}. Applied to the LEEM electron beam, it reads $c=\frac{1}{\Sigma}(1-\frac{I(\theta)}{I_0})$, where $\Sigma$ is the effective electron-adatom cross section and $I_0$ is the specular intensity with zero coverage. By comparing the LEEM $I(\varepsilon)$ curves of a sample before and after the deposition, a maximum of sensitivity to the deposited species can be determined. % as  [Fig. \ref{fig:methods}(b)]
As shown in the the inset of Fig. S2, a significant reflectivity change induced by the Ag-2DAG deposition is observed for an incident electron energy of \unit[24]{eV}. Using the slope at the origin  in Fig. S2 to determine $\Sigma$ (black line), the typical intensity variation upon Ag deposition on the Si(111)$\sqrt{3}\times\sqrt{3}$-Ag gives $c=0.43\times(1-\frac{I(\theta)}{I_0})$ and $\Sigma=\unit[29\pm3]{\mathring{A}^2}$, assuming a Ag-2DAG adsorption site density equal to that of the Si(111)$\sqrt{3}\times\sqrt{3}$-Ag (\emph{i.e.} \unit[7.83$\times$10$^{14}$]{atoms/cm$^{2}$} \cite{nak97}). This translate into a concentration measurement uncertainty of \unit[$\simeq$10]{\%}. The measured values of $\Sigma$ are in quantitative agreement with previous study by LEEM on Ag/W(100) \cite{fig06} and also close to the value of \unit[47]{$\mathring{\mathrm{A}}^2$} determined from He scattering measurements on Ag/Pd(100) \cite{far98}. 

%\begin{figure}[h!]
%	\centering
%	\includegraphics[width=90mm]{Fig8b}\caption{\label{fig:methodsConc} Surface reflectivity change induced by a Ag-2DAG deposition at RT (red curve) and the deduced Ag-2DAG concentration monitoring (blue curve, see text for details). The dashed dark line is a linear fit used to derive the value of $\Sigma$ \cite{far98}. The inset shows the reflectivity curves of a $\sqrt{3}\times\sqrt{3}$-Ag surface before and after a \unit[0.09]{ML} Ag-deposition at \unit[220]{K}. A clear sensitivity to the deposition is observed at \unit[24]{eV}.}
%	%2015-01-16
%	%Inset 2014-05-14
%\end{figure}

The adspecies concentration monitoring allows to identify different Ag-deposition related regimes. First, the Ag-2DAG concentration increases linearly with time or equivalently with the amount of deposited Ag until it reaches a critical concentration ($c_{\mathrm{nucl}}$) where 3D islands begin to grow on an unmodified $\sqrt{3}$-Ag surface as expected for a Stranski-Krastanov growth mode. 
%%[Fig. \ref{fig:conc}(b)]
%. % as illustrated in Fig. \ref{fig:morph}(a).    
After the 3D phase nucleation, the Ag-2DAG concentration decreases and reaches a stationary state ($c_{\mathrm{stat}}$) where Ag adatom deposit, surface diffusion and consumption by the 3D growing phase are in a dynamical equilibrium. %This typical deposition
%highlights that a supersaturation of the Ag-2DAG is required to nucleate the 3D growing phase.%. 
2DAG concentration maps can also be determined. The $c(\frac{I}{I_\mathrm{0}})$ relationship is computed pixel-by-pixel (instead of micron-sized homogeneous areas) between an image obtained at \unit[24]{eV} for a given coverage and a reference acquired before any deposition.

\section*{References}

%\bibliographystyle{elsarticle-num}
%\bibliography{AgOnSi111}

%\section*{Acknowledgements}

%Acknowledgements should be brief, and should not include thanks to anonymous referees and editors, or effusive comments. Grant or contribution numbers may be acknowledged.

%\section*{Author contributions statement}
%
%F.C. conceived and carried out the experiments. F.C. analysed the results and all authors took part in discussing the results. F.C. wrote the manuscript text with input from all the authors. All authors reviewed the manuscript. 
%
%\section*{Additional information}
%
%The authors declare no competing financial interests.

\end{document}